\documentclass[aps,preprint,superscriptaddress,amssymb,amsmath,nofootinbib]{revtex4}

\usepackage{graphicx}
\usepackage{ifpdf}
\pdfoutput=1 

\def\lqcd{\Lambda_{\rm QCD}}
\def\OMIT#1{{}}

\def\xbar{\bar x}
\def\pplus{\hat{p}_+}
\def\eqn#1{Eq.~(\ref{#1})}
\def\digamma{\psi}

\def\hgammaU{\widehat{\Gamma}_u^p}
\def\hgammaS{\widehat{\Gamma}_{77}^p}
\def\tgammaU{\widetilde{\Gamma}_u^p}
\def\tgammaS{\widetilde{\Gamma}_{77}^p}
\def\wdp{W(\Delta, P_\gamma)}
\def\msbar{\overline{\rm MS}}

\def\fig#1{Fig.\ \ref{#1}}

\def\vereq#1#2{\lower3pt\vbox{\baselineskip1pt\lineskip1pt
     \ialign{\nonumber \\$#1\hfill##\hfil\nonumber \\$\crcr#2\crcr\sim\crcr}}}

\makeatletter
\def\fmslash{\@ifnextchar[{\fmsl@sh}{\fmsl@sh[0mu]}}
\def\fmsl@sh[#1]#2{%
   \mathchoice
     {\@fmsl@sh\displaystyle{#1}{#2}}%
     {\@fmsl@sh\textstyle{#1}{#2}}%
     {\@fmsl@sh\scriptstyle{#1}{#2}}%
     {\@fmsl@sh\scriptscriptstyle{#1}{#2}}}
\def\@fmsl@sh#1#2#3{\m@th\ooalign{$\hfil#1\mkern#2/\hfil$\crcr$#1#3$}}

\makeatother
\begin{document}
\ifpdf
\DeclareGraphicsExtensions{.pdf, .jpg}
\else
\DeclareGraphicsExtensions{.eps, .jpg}
\fi


\title{High order perturbative corrections to the determination of $|V_{ub}|$
from the $P_+$ spectrum in $B\rightarrow X_{u}\ell\bar\nu_\ell$}
\author{Francisco Campanario}
\email{francam@particle.uni-karlsruhe.de}
\affiliation{Departament de F\'{\i}sica Te\`orica and IFIC, 
Universitat de Val\`encia - CSIC, E-46100 Burjassot, Val\`encia, Spain}
\affiliation{Institut f\"ur Theoretische Physik, 
        Universit\"at Karlsruhe, P.O.\ Box 6980, 76128 Karlsruhe, Germany}
\author{Michael Luke}
\email{luke@physics.utoronto.ca}
\affiliation{Department of Physics, University of Toronto, 
     60 St.\ George Street, Toronto, Ontario, Canada M5S 1A7}
\author{Saba Zuberi}
\email{szuberi@physics.utoronto.ca}
\affiliation{Department of Physics, University of Toronto, 
     60 St.\ George Street, Toronto, Ontario, Canada M5S 1A7}\date{\today}

\begin{abstract}
We investigate the behaviour of the perturbative relation between the photon energy spectrum in $B\to X_s\gamma$ and the hadronic $P_+$ spectrum in semileptonic $B\to X_u\ell\bar\nu$ decay at high orders in perturbation theory in the ``large-$\beta_0$" limit, in which only terms of order $\alpha_s^n\beta_0^{n-1}$ are retained.   The leading renormalon in the weight function $W(\Delta,P_{\gamma})$ relating the two spectra is confirmed to be at $u=1/2$, corresponding to nonperturbative corrections at $O(\Lambda_{QCD}/m_b)$.   We show that the $P_\gamma$ dependent pieces of the weight function have no infrared renormalons in this limit, and so the factorial growth in perturbation theory arises solely from the constant terms.   We find no numerical enhancement of leading logarithms, suggesting that fixed-order perturbation theory is more appropriate than a leading-log resummation for the extraction of $|V_{ub}|$.  The importance of various terms in the expansion of the weight function is studied using a model for the $B\rightarrow X_{s}\gamma$ photon spectrum.  Our analysis suggests that higher order perturbative corrections do not introduce a significant uncertainty in the extraction of $|V_{ub}|$.
\end{abstract}

\preprint{FTUV-08-1022}
\preprint{KA--TP--07--2008}

\maketitle

\section{Introduction}

The total rate for the decay ${B}\rightarrow X_{u}\ell\bar\nu$ provides  a  theoretically clean determination of the magnitude of the CKM matrix element $\vert V_{ub}\vert$ as a double expansion in powers of $\alpha_s(m_b)$ and $\lqcd/m_b$ \cite{OPE}.   However, to eliminate the background from $B\to X_c$ decays, strong cuts on the final state phase space are required, which can complicate the theoretical analysis.   The kinematic regions in which cuts on the charged lepton energy $E_\ell$, hadronic invariant mass $m_X$ \cite{mXcut} and hadronic light-cone momentum $P_+=E_X-|\vec P_X|$ (where $E_X$ and $\vec P_X$ are the energy and three-momentum of the final state hadrons) \cite{p+cut} are strong enough to eliminate the charm background all correspond to the so-called shape function regime, in which the local OPE for the partial rate breaks down \cite{shapefunct,shapefunct2}.  However, in this region an expansion of the partial rate in powers of $\lqcd/m_b$ in terms of non-local operators is still possible, and the matrix element of the leading nonlocal operator can be measured in $B\to X_s\gamma$ decay.  At leading order in $\lqcd/m_b$, we can write
\begin{equation}
d\Gamma_i=\int C_i(\omega) f(\omega)+O\left({\lqcd\over m_b}\right)
\end{equation}
where $i$ labels the decay, $C_i(\omega)$ is perturbatively calculable, and the shape function $f(\omega)$ is nonperturbative, but universal in inclusive $B$ decays.\footnote{$C(\omega)$ can be further factorized into ``hard" and ``jet" functions; however, for our purposes we will not make this decomposition.}   It is convenient to eliminate the shape function altogether, and express integrated rates directly in terms of one another \cite{shapefunct2,Leibovich,HLLP+}.  For example, we can write
\begin{equation}\label{wdefinition}
\int_0^\Delta dP_+{d\Gamma_u\over dP_+}\propto \int_0^\Delta dP_\gamma W(\Delta, P_\gamma) {d\Gamma_s\over dP_\gamma}+O\left({\lqcd\over m_b}\right)
\end{equation}
where $P_\gamma\equiv m_B-2 E_\gamma$, $E_\gamma$ is the photon energy and $\Delta\sim O(\lqcd)$.  This defines the weight function $W(\Delta, P_\gamma)$, which can be calculated in perturbation theory. The $O(\lqcd/m_b)$ power corrections have been extensively discussed in the literature \cite{subleadingshape1,subleadingshape2,neub2loop,Stewart:2004a,Beneke:2004,Lee:2006wn}, and have typically been estimated to be below the $10\%$ level for $|V_{ub}|$ \cite{neub2loop,Stewart:2004a,Beneke:2004}, although it has been argued that subleading four-quark operators may introduce significant uncertainties \cite{Stewart:2004a}. 

The weight function $W(\Delta, P_\gamma)$ has been calculated in fixed-order perturbation theory to $O(\alpha_s^2\beta_0)$ \cite{HLLP+}. It is also known to next-to-leading-log order, $O(\alpha_s^n \log^{n-1} m_b/\mu_i)$, where $\mu_i\sim \sqrt{\lqcd m_b}$ is the typical invariant mass of the final state \cite{neub2loop}, generalized in  \cite{Lange}.   
It was shown in \cite{HLLP+} that the $O(\alpha_s^2\beta_0)$ corrections to $W(\Delta, P_\gamma)$ are substantial, and the same order as the $O(\alpha_s)$ corrections.  Given the size of these corrections, it is important to study the convergence of the perturbative expansion.

In this paper we examine the behaviour of $W(\Delta, P_\gamma)$ at higher fixed orders in perturbation theory.  We work in the framework of the ``large-$\beta_0$" expansion, in which we calculate all terms of order $\alpha_s^n\beta_0^{n-1}$ \cite{BB,Ball}.  While there is no limit of QCD in which these terms formally dominate, this class of terms allows us to examine the asymptotic nature of perturbation theory, as well as giving an estimate for the size of perturbative corrections.  We discuss the significance of these terms for the extraction of $|V_{ub}|$.

\section{Borel Transformed Spectra and the Weight Function}

Since QCD has an asymptotic perturbative expansion, it is convenient to study the Borel transformed series $B[\widetilde{R}](u)$ of a quantity $\widetilde{R}$, where
\begin{equation}
\widetilde{R}=R-R_{\text{tree}}=\sum_{n=0}^\infty  r_n \alpha_s^{n+1} 
\end{equation}
and
\begin{equation}\label{defborel}
B[\widetilde{R}](u)=\sum_{n=0}^\infty {r_n\over n!} u^n.
\end{equation}
The expansion for $B[\widetilde{R}](u)$ has better convergence properties than the original expansion.
$B[\widetilde{R}](u)$ can in turn be used as a generating function for the coefficients $r_{n}$
\begin{equation}\label{rn}
r_{n}=\frac{d^{n}}{du^{n}}B[\widetilde{R}](u)|_{u=0}
\end{equation}
while the original expression $R$ can be
recovered from the Borel transform $B[\widetilde{R}](u)$ by the inverse Borel
transform
\begin{equation}\label{inverse}
  R= R_{\text{tree}}+\int_0^\infty du\ e^{-u/\alpha_s}\ B[\widetilde{R}](u).
\end{equation}
Singularities in $B[\widetilde{R}](u)$ along the positive real $u$ axis make
the inverse Borel transform ill-defined.  These are referred to as
infrared renormalons \cite{thooft}, factorially growing contributions to the coefficients
of the perturbative series, which lead to ambiguities of order 
$(\Lambda_{\text{QCD}}/m_b)^n$.
In physical quantities these
ambiguities are compensated by corresponding ambiguities in the
definition of higher-dimensional nonperturbative matrix elements in the operator product
expansion of order $\Lambda_{\text{QCD}}^n$, which render the physical quantity
unambigious.\footnote{Although the renormalon cancellation has only been explicitly shown in some cases in the large-$\beta_0$ limit,  it is assumed to hold away from this limit.}  

The Borel transform \eqn{defborel}, in the large-$\beta_0$ limit, may be determined from the order $\alpha_{s}$ term, $r_{0}$, with finite gluon mass following \cite{Ball}: 
\begin{eqnarray} \label{borelTransf}
B[\widetilde{R}](u)&=&-\frac{ \sin \pi u}{\pi u} e^{5 u/3} \int_0^{\infty }
\left(\frac{\lambda^{2}}{m_b^2}\right)^{-u}
\left(\frac{dr_{0}}{d\lambda^2} - \frac{r_{\infty }}{\lambda^{2}}
  \Theta(\lambda^{2}-m_b^2 e^{5 u/3})\right)\, d\lambda^{2} \nonumber \nonumber \\
&&+\frac{1}{u}\left( \widehat{G}_{0}(u)-\frac{ \sin \pi u}{\pi u} r_{\infty}\right).
\end{eqnarray}
Here $\lambda$ is the gluon mass and $r_{\infty}$ is a constant. We have
used the $\msbar$ scheme with the renormalization scale $\mu$ set
to the pole mass, $m_b$. The terms $\widehat{G}_0(u)/u$ and $r_\infty$ arise
from the renormalization of the graphs involved.

The weight function $W(\Delta, P_\gamma)$ is defined through the relation between the integrated $B\to X_s\gamma$ photon energy spectrum and $B\to X_u\ell\bar\nu$ charged lepton spectrum,
\begin{eqnarray} \label{weightRelation}
\Gamma_u(\Delta)&\equiv&\int_0^\Delta dP_{+}\frac{d\Gamma _{u}}{dP_{+}}
= \frac{|V_{ub}|^{2}}{|V_{tb}V_{ts}^{\ast }|^{2}} \frac{\pi}{6 \alpha_{em} C_{7}^{\rm eff}(m_b)^{2}} \frac{m_B^2}{\overline m_b(m_b)^2}\int_0^\Delta dP_{\gamma }W(\Delta ,P_{\gamma })\frac{d\Gamma _{77}
}{dP_{\gamma }}\nonumber\\
&&+O\left( \frac{\Lambda _{QCD}}{m_b}\right)
\end{eqnarray}
where $\Delta \sim \Lambda _{QCD}$ in the shape function region, and the normalization is the same as that used in \cite{HLLP+}.  Other definitions of $W$ are possible, such as that used in \cite{neub2loop}.  As in \cite{HLLP+}, we concentrate on the contribution to the $B\to X_s\gamma$ spectrum arising from the operator $O_{7}=(e/16 \pi^2) m_b \bar{s_{L}} \sigma^{\mu \nu} F_{\mu \nu} b_{R}$.  While other operators also contribute to the spectrum, for the purposes of studying the convergence of the series and estimating the uncertainties from higher order terms in perturbation theory we will neglect their contribution and the mixing of these with $O_7$.   The factor of $m_B^2/\bar m_b^2$ pulled out of the relation arises naturally, and improves the behaviour of perturbation theory for $W(\Delta, P_\gamma)$\cite{HLLP+}.

Defining the partonic partial rates
\begin{eqnarray} 
\frac{1}{\Gamma_\gamma} \frac{d\Gamma _{77}}{d \bar{x}}&=&\delta(\xbar)+g(\xbar)\nonumber\\
\frac{1}{\Gamma_u} \frac{d\Gamma _{u}}{d \hat{p}_{+}} &=&\delta(\pplus)+h(\pplus)
\end{eqnarray}
where $\Gamma_{\gamma}=G_{F}^2 |V_{tb}V_{ts}^{\ast }|^{2} \alpha_{em} m_b^3[\overline m_b(m_b) C_{7}^{\rm eff}(m_b)]^2/(32 \pi^4)$ and $\Gamma_u=G_F^2 |V_{ub}|^2 m_b^5/(192 \pi^3)$ are the leading order widths.  The partonic variables 
\begin{equation} \xbar\equiv 1-2 E_{\gamma}/m_b, \qquad \pplus\equiv (v-q/m_b)\cdot n \end{equation}
are related to the hadronic variables by
\begin{equation}
P_{\gamma} \equiv m_B-2 E_{\gamma}=m_b \bar{x}+\Lambda,\qquad P_+\equiv E_X-|\vec P_X|=m_b \pplus +\Lambda
\end{equation}
where $\Lambda \equiv m_B-m_b$, $q$ is the momentum of the lepton-neutrino pair, $n$ is a light-like four vector in the $-\vec{q}$ direction and $v$ is the four-velocity of the $B$ meson.  
Convoluting the partonic rate with the shape function to obtain the hadronic rates, we find
\begin{equation}\label{Wextract}
W(\Delta, P_\gamma)=1+\int_0^{\Delta-P_\gamma}\left(h(p)-g(p)\right) dp-\int_0^{\Delta-P_\gamma} g(p)\left[ W(\Delta, p+P_\gamma)-1\right] dp
\end{equation}
where the partonic spectra are expanded to leading order in $\bar{x}$ and $\hat{p}_{+}$ respectively since in the shape function region they are of $O(\Lambda_{QCD}/m_b)$.  

Since $g(p)$ and $h(p)$ are $O(\alpha_s)$, \eqn{Wextract} may be solved iteratively for $W(\Delta, P_\gamma)$.  For the purposes of this paper, we are only interested in terms of $O(\alpha_s^n\beta_0^{n-1})$, for which the last term in \eqn{Wextract} does not contribute; therefore, we can write
\begin{equation} 
W(\Delta, P_\gamma)=\hgammaU(\Delta-P_\gamma)-\hgammaS(\Delta-P_\gamma)+O(\alpha_s^n \beta_0^{n-2})
\end{equation}
where we have defined the integrated partonic rates calculated in perturbation theory,
\begin{equation}
\hgammaS(\Delta)=\frac{1}{\Gamma_\gamma} \int_{0}^{\Delta} \frac{d\Gamma _{77}}{d\xbar}d\xbar
\end{equation}
and
\begin{equation}
\hgammaU(\Delta)=\frac{1}{\Gamma_u} \int_0^\Delta \frac{d\Gamma _u}{d\pplus}d\pplus.
\end{equation}
The corresponding quantities $\widetilde{W}$, $\tgammaS$ and $\tgammaU$ are defined by subtracting the tree
level contribution.

Calculating the parton level photon spectrum with a massive gluon is straightforward, and was done in \cite{HLLP+}. Integrating the rate with a massive gluon over the endpoint region and performing the integral \eqn{borelTransf}, we find the Borel transform of the integrated partonic rate: 

\begin{eqnarray}\label{borelbtos}
B[\tgammaS(\Delta)](u) &=&e^{5 u/3} \left(\frac{2 (u-1)}{u^2} \left(\frac{\Delta}{m_b}\right) ^{-2 u}-\left(\frac{2}{u-1}-\frac{3}{u}-\frac{4}{u^2}+\frac{1}{u-2}\right) \frac{ \sin \pi  u}{\pi  u} \left(\frac{\Delta}{m_b}\right)^{-u}\right.\nonumber \\
&&+\left.\frac{2 \sin \pi  u}{\pi  u^2}+\frac{(1+u) \left(3 u^2-2 u-2\right) \Gamma (u)^2 }{(u-2) (u-1)u\Gamma (2 u)\cos \pi  u}\right)\nonumber \\
&&+\frac{1}{u}\left(\widehat{G}_0(u)-\frac{ 2 e^{5 u/3} \sin \pi  u}{\pi  u}\right).
\end{eqnarray}
Since the operator $O_{7}$ requires renormalization, the last line arises from the $\msbar$ counterterm.
$\widehat{G}_{0}(u)$ is given by
\begin{equation}\label{subtract}
\widehat{G}_{0}(u) = \sum_{n=0}^{\infty} \frac{g_{n}}{n!} u^n
\end{equation}
and $g_{n}$ are the coefficients of the expansion of $G_{0}(u)$  \cite{Ball}
\begin{equation}\label{subtract2}
G_{0}(u) = \sum_{m=0}^{\infty} g_{m} u^m = \frac{2 (2 u+1) \Gamma(4+2 u)}{3 (u+2) (u+1) \Gamma(2+u)^2}\frac{\sin\pi u}{\pi u}.
\end{equation}
The Borel transform of the differential photon spectrum away from the $\bar x=0$ endpoint was calculated in \cite{GardiAndersen}.   Integrating this result from $\xbar=0$ to $\xbar=\Delta$ reproduces the $\Delta$ dependent terms of our result, \eqn{borelbtos}.  (The $\Delta$-independent terms depend on the virtual contribution and cannot be directly compared against \cite{GardiAndersen}).

The calculation of the Borel transform of the semileptonic partial rate $\hgammaU(\Delta)$ is significantly more involved than for $B\to X_s\gamma$.  The Borel transform of the triple-differential ${B}\rightarrow X_{u}\ell\bar\nu$ spectrum was calculated in \cite{Gambino}.   Rather than integrate this result over the appropriate phase space, we instead calculated the integrated rate $\Gamma_u(\Delta)$ for a massive gluon, and then performed the integral (\ref{borelTransf}).  The result has the comparatively simple form
\begin{eqnarray}\label{borelbtou}
B[\tgammaU(\Delta)](u) &=& e^{5 u/3}\left( \frac{2 (u-1)}{u^2}\left(\frac{\Delta}{m_b}\right)^{-2 u}\right.\nonumber\\
&&+\left(\frac{5}{3 (u-3)}-\frac{2}{u-2}-\frac{5}{3 (u-1)}+\frac{7}{3 u}+\frac{2}{u^2}-\frac{1}{3 (u-4)}\right) \frac{2 \sin \pi u}{\pi u} \left(\frac{\Delta}{m_b}\right)^{-u}\nonumber\\
&&+\frac{\Gamma(u)^2 }{(u-4)(u-2)(u-1) u \Gamma(2 u)\cos\pi u}\left(\frac{1}{3}(9u^4-103u^3-62u^2+38u+24)\right.\nonumber\\
&&\left. \left. -\vphantom{1\over 3}16 u (1+u)(2 u-1)\left({\pi \over\sin 2\pi u} +\digamma(u)-\digamma(2 u) \right)\right) \right)  
\end{eqnarray}
where $\digamma(u)=\Gamma^\prime(u)/\Gamma(u)$ is the digamma function.  

The Borel transformed weight function is given by the difference between Eq.~(\ref{borelbtou}) and Eq.~(\ref{borelbtos}).  Note that the terms proportional to $(\Delta/m_b)^{-2 u}/u^2$ and $ (\Delta/m_b)^{-u} \sin\pi u/u^3$, which generate the $\alpha_s^n  \ln^{n+1}(\Delta/m_b)$ logs, cancel in the difference. This reflects the universality of the leading Sudakov logs. We can resum this contribution by evaluating the inverse Borel transform, \eqn{inverse}. However the result does not exponentiate because higher powers of logs, up to $\alpha_s^n \ln^{2n}$ double logs, are not included since they are suppressed in $\beta_0$. The resummed $\alpha_s^n  \ln^{n+1}(\Delta/m_b)$ logs from \eqn{btosalphacoeff} and \eqn{btoualphacoeff} are given by

\begin{eqnarray}\label{sudakov}
\frac{C_F}{\beta_{0}}\int_{0}^{\infty} du\ e^{-\frac{4 \pi u}{\alpha_{s} \beta_{0}}} \frac{2}{u^2} \left(-\left(\Delta\over m_b\right)^{-2 u}+2 \left(\Delta\over m_b\right)^{-u}-1\right)\nonumber\\
= \frac{C_F}{\beta_0} \left(4 \ln \frac{\Delta}{m_b} \ln \frac {1+a}{1+ 2 a} + \frac{8\pi}{\alpha_s \beta_0}\ln \frac{ \left({1+a} \right)^2}{1+2 a} \right)
\end{eqnarray}
where $a \equiv {\alpha_s(m_b)\beta_0\over 4\pi} \ln {\Delta\over m_b}$.

The final result for the Borel transformed weight function is

\begin{eqnarray}\label{borelW}
B[\widetilde{W}(\Delta,P_{\gamma})](u) &=&e^{5 u/3} \left( \frac{2 \sin\pi u}{\pi u^2} \left( \frac{(u-5)(3 u-4)}{(u-4)(u-3)(u-2)(u-1)} \left(\frac{\Delta-P_{\gamma}}{m_b}\right)^{-u}-1\right)\right.\nonumber\\
&&\hskip-1in-\frac{\Gamma(u)^2}{\Gamma(2 u)(u-4)(u-2)(u-1) \cos \pi u}\left(\vphantom{2\over 3}16 (u+1)(2 u-1)\left({\pi \over \sin 2 \pi u} +\digamma(u)-\digamma(2 u) \right) \right.\nonumber\\
&&\hskip-1in\left.\left.+\frac{2}{3}(5 u+2)(7u+1) \right) \right)-\frac{1}{u}\left(\widehat{G}_0(u)-\frac{ 2 e^{5 u/3} \sin \pi  u}{\pi  u}\right)
\end{eqnarray}
where $\widehat{G}_0(u)$ is obtained from Eq.~(\ref{subtract}) and Eq.~(\ref{subtract2}).\eqn{borelW} is the main result of this paper.

The Borel transforms can be used to generate the $O(\alpha_s^n\beta_0^{n-1})$ terms in the perturbative expansion via the relation \eqn{rn}. Writing
\begin{eqnarray}
\hgammaS(\Delta)&=&1+\sum_{i=1}^\infty C^s_n(\Delta) {\alpha_s(m_b)^n\beta_0^{n-1} C_F\over (4\pi)^n}\nonumber\\
\hgammaU(\Delta)&=&1+\sum_{i=1}^\infty C^u_n(\Delta) {\alpha_s(m_b)^n\beta_0^{n-1} C_F\over (4\pi)^n}\nonumber\\
W(\Delta,P_\gamma)&=&1+\sum_{i=1}^\infty W_n(\Delta, P_\gamma) {\alpha_s(m_b)^n\beta_0^{n-1} C_F\over (4\pi)^n}
\end{eqnarray}
we can easily find the coefficients $C^s_n(\Delta)$, $C^u_n(\Delta)$ and $W_n(\Delta, P_\gamma)$ to any order. The coefficients are given up to $n=5$ in Appendix \ref{appendixa}.  

The leading log (LL) and next-to-leading log (NLL) terms in \eqn{wAlphaExp} are contained within the renormalization group resummed NLL result in soft-collinear effective theory (SCET), $W(\Delta,P_{\gamma})^{\rm NLL}_{\rm SCET}$, obtained from \cite{p+cut,Neubert,BauerMan,neub2loop}.  The SCET result sums logs of $\mu_i^2/m_b^2$, where $\mu_i^2 \sim O(\Lambda_{QCD} m_b) $. In the Appendix  \ref{appendixb} we verify that the leading $\beta_0$ terms agree with \eqn{wAlphaSCET} by expanding in $\alpha_s(m_b)$ and then expanding logs of $\mu_i^2/m_b^2$ and $\mu_i^2/(m_b (\Delta-P_\gamma))$. Our results also agree with those in \cite{HLLP+,MLphoton,ARF1}.

\section{Results and Discussion}

\subsection{Renormalons and Borel Resummation}

The leading renormalon ambiguity in both the photon and semileptonic spectra is $O(\Lambda_{QCD}/m_b)$ due to the pole at $u=1/2$ in $B[\tgammaU(\Delta)](u)$ and $B[\tgammaS(\Delta)](u)$. The divergence does not cancel between the spectra and gives rise to a pole at $u=1/2$ in the Borel transformed weight function. This is consistent with the presence of nonperturbative corrections to $W(\Delta,P_{\gamma})$ at $O(\Lambda_{QCD}/m_b)$ due to subleading shape functions \cite{subleadingshape1}.   
  
The Borel transform of the weight function can be written in terms of $\Delta-P_{\gamma}$ independent and dependent pieces, $B[\widetilde{W}_0](u)$ and  $B[\widetilde{W}_1](u)$ respectively,
\begin{eqnarray}\label{borelW0}
B[\widetilde{W}_0](u) &=&e^{5 u/3} \left( \frac{2 \sin\pi u}{\pi u^2} \left( \frac{(u-5)(3 u-4)}{(u-4)(u-3)(u-2)(u-1)}-1\right)\right.\nonumber\\
&&\hskip-.6in-\frac{\Gamma(u)^2 }{\Gamma(2 u)(u-4)(u-2)(u-1)\cos \pi u}\left(\vphantom{2\over 3}16 (u+1)(2 u-1)\left({\pi \over \sin 2 \pi u} +\digamma(u)-\digamma(2 u) \right) \right.\nonumber\\
&&\hskip-,6in\left.\left.+\frac{2}{3}(5 u+2)(7u+1) \right) \right)-\frac{1}{u}\left(\widehat{G}_0(u)-\frac{ 2 e^{5 u/3} \sin \pi  u}{\pi  u}\right)
\end{eqnarray}

\begin{eqnarray}\label{borelW1}
B[\widetilde{W}_1(\Delta, P_\gamma)](u) &=&e^{5 u/3}  \frac{2 \sin\pi u}{\pi u^2} \frac{(u-5)(3 u-4)}{(u-4)(u-3)(u-2)(u-1)} \left(\left(\frac{\Delta-P_{\gamma}}{m_b}\right)^{-u}-1\right)
\end{eqnarray}
where we have defined $B[\widetilde{W}_0](u)$ and $B[\widetilde{W}_1](u)$ such that they are finite as $u\to0$. Note that $B[\widetilde{W}_1](u)$ has no singularities for positive $u$. Therefore the inverse Borel transform of \eqn{borelW1}, $\widetilde{W}_1$, is well defined and unambiguously resums logarithms of $(\Delta-P_{\gamma})/m_b$. This tells us that the poor behavior in the perturbative expansion of the weight function is entirely due to the constant terms, $\widetilde{W}_0$, which are generated by $B[\widetilde{W}_0](u)$.

The relevant quantity in determining $|V_{ub}|$ is the weight function convoluted with the $B\rightarrow X_{s}\gamma$ photon spectrum, as in \eqn{weightRelation}. It is interesting to note that the integrated quantity can have a renormalon ambiguity that is not present in the weight function. In order to illustrate this we calculate the Borel transform of $\widetilde{W}_1$, which is renormalon free, convoluted with a simple model of the normalized $B\rightarrow X_{s}\gamma$ spectrum,
\begin{equation}\label{normPhoton}
\frac{1}{\Gamma_{\gamma}} \frac{d\Gamma_{s}}{d P_{\gamma}} = \frac{b^{b}}{\Gamma(b) \Lambda^{b}} P_{\gamma}^{b-1} e^{-\frac{b P_{\gamma}}{\Lambda}}
\end{equation}
where $b=2.5$ and $\Lambda=0.77$ GeV  \cite{neub2loop}. This is straightforward to obtain from \eqn{borelW1}:
\begin{eqnarray}\label{borelWint}
B\left[\int_0^\Delta dP_\gamma \widetilde{W}_1\frac{1}{\Gamma_{\gamma}} \frac{d\Gamma_{s}}{d P_{\gamma}}\right](u) &=&e^{5 u/3}  \frac{2 \sin\pi u}{\pi u^2} \frac{(u-5)(3 u-4)}{(u-4)(u-3)(u-2)(u-1)}\left(\frac{\Gamma\left(b,{b \Delta \over \Lambda}\right)}{\Gamma(b)}\right.\nonumber\\
&&\hskip-1in \left.+\left(-1+\left({b \Delta \over\Lambda}\right)^b \left({\Delta\over m_b}\right)^{-u} \frac{\Gamma(1-u)}{\Gamma(1-u+b)} \phantom{}_1F_1\left(b;1-u+b;{-b \Delta\over\Lambda}\right)\right)\right)
\end{eqnarray}
where $\Gamma(a,z)=\int_z^\infty t^{a-1} e^{-t} dt$ is the incomplete Gamma function.
The $\Gamma(1-u)$ term in \eqn{borelWint} gives rise to a pole at $u=1$, which corresponds to an order $O((\Lambda_{QCD}/m_b)^2)$ ambiguity in the integrated quantity. This arises because higher order terms in the perturbative expansion of $\widetilde{W}_1$ have more powers of $\ln (\Delta-P_\gamma)/m_b$ and therefore are more singular near the end point. However since the renormalon in $B[\widetilde{W}_0](u)$ is at $u=1/2$, the factorial growth in the integrated quantity is dominated by the constant terms in the weight function rather than the logarithms.

It is amusing to notice that if the $\alpha_s^n \ln(\Delta/m_b)^{n+1}$ Sudakov logs did not cancel between $\hgammaU(\Delta)$ and $\hgammaS(\Delta)$ these would give rise to a $\left((\Delta-P_\gamma)/ m_b\right)^{-2 u}$ term in the Borel transform of the weight function. When integrated over $P_\gamma$ with \eqn{normPhoton} this would lead to a pole at $u=1/2$, the same order as the renormalon in $B[\widetilde{W}_0](u)$. 

Since $B[\widetilde{W_1}](u)$ has no poles in $u$, the inverse Borel transform of \eqn{borelW1} is well-defined.  We may therefore use \eqn{borelW1} to sum all terms containing powers of $\ln((\Delta-P_\gamma)/m_b)$ (terms of order $\alpha_s^n\beta_0^{n-1}\log^{n-m}(\Delta-P_\gamma)/m_b$, for $n=1$ to infinity and $m=0$ to $n-1$).   While we were unable to obtain a closed-form result for this quantity, by expanding \eqn{borelW1} in powers of $u$ it is straightforward to sum all terms of order  $\alpha_s^n\beta_0^{n-1}\log^{n-m}(\Delta-P_\gamma)/m_b$, for $n=m+1$ to infinity and for fixed $m\geq 0$, by evaluating the inverse Borel transform

\begin{eqnarray}\label{inverseBorel}
W(\Delta,P_{\gamma})_{\rm resummed}
&=&\frac{C_F}{\beta_{0}}\int_{0}^{\infty} du\ e^{-\frac{4 \pi u}{\alpha_{s} \beta_{0}}} C_{m-1} u^{m-1} \left(\left(\frac{\Delta-P_{\gamma}}{m_b}\right)^{-u}-1\right)\nonumber\\
&=&\frac{C_F C_{m-1}}{\beta_{0}} \Gamma(m) \left(\frac{\alpha_{s} \beta_{0}}{4 \pi}\right)^{m} \left( \left(1+\frac{\alpha_{s} \beta_{0}}{4 \pi}\ln\frac{\Delta-P_{\gamma}}{m_b} \right)^{-m}-1 \right)
\end{eqnarray}
where $C_{m-1}$ is the coefficient of the $u^{m-1} \left(\frac{\Delta-P_{\gamma}}{m_b}\right)^{-u}$ term in \eqn{borelW1}, and the second line follows for $m>0.$  The constant non-logarithmic terms in the weight function are not included in \eqn{inverseBorel}, as they arise from $\widetilde {W}_0$, but may be obtained from the expansion \eqn{wAlphaExp} to give the full resummed logarithmic result. At leading-log (LL), $m=0$, we find
\begin{equation}\label{wLLbeta0}
\begin{array}{l}
W(\Delta,P_{\gamma})_{\beta_{0}}^{LL}=1-\frac{5 C_F}{3 \beta_{0}}\ln\left(\frac{\alpha_{s}(m_b) \beta_{0}}{4 \pi} \ln\frac{\Delta-P_{\gamma}}{m_b}+1\right).
\end{array}
\end{equation}
We explicitly show the NLL, the next-to-next-to-leading logarithmic (NNLL) $\alpha_{s}^{n} \beta_{0}^{n-1} \ln^{n-2}$ and next-to-next-to-next-to-leading logarithmic (NNNLL) $\alpha_{s}^{n} \beta_{0}^{n-1} \ln^{n-3}$ terms below:
\begin{eqnarray}\label{wNLLbeta0}
W(\Delta,P_{\gamma}) _{\beta_{0}} ^{NLL}&=&\frac{\alpha_{s}(m_b) C_F}{4 \pi} \left[{14\over 3}\left(\frac{1}{1+b}-1\right)+ \left(\frac{167}{36}-\frac{2 \pi^2}{3}\right)\right]
\nonumber\\
W(\Delta,P_{\gamma}) _{\beta_{0}} ^{NNLL}&=& \frac{\alpha_{s}(m_b)^{2} \beta_{0}C_F}{(4 \pi)^{2}} \left[\left(\frac{1559}{216}-\frac{5 \pi^{2}}{18}\right) \left( \frac{1}{\left(1+b\right)^{2}}-1\right)\right.\nonumber \\
&& +\left. \left(\frac{3857}{144}-\frac{16 \pi^2}{9}-12 \zeta(3)\right)\right]
\nonumber\\
W(\Delta,P_{\gamma}) _{\beta_{0}} ^{NNNLL}&=&  \frac{\alpha_{s}(m_b)^{3} \beta_{0}^{2}C_F}{(4 \pi)^{3}}\left[ \left(\frac{65545}{3888}-\frac{14 \pi^{2}}{9}\right)\left(\frac{1}{(1+b)^{3}}-1\right)\right.\nonumber \\
&&+\left. \left(\frac{90043}{864}-\frac{13 \pi^2}{108}-\frac{16 \pi^4}{15}-\frac{166 \zeta(3)}{3}\right)\right].
\end{eqnarray}
where $b\equiv \frac{\alpha_{s}(m_b) \beta_{0}}{4 \pi} \ln\frac{\Delta-P_{\gamma}}{m_b}$.
These results provide a useful check of our calculation, as they may be compared with the corresponding resummed expressions in SCET, obtained from \cite{p+cut,Neubert,BauerMan,neub2loop}. Setting $\mu_i=\sqrt{m_b (\Delta-P_{\gamma})}$, we verify that the resummed LL and NLL contributions in the large $\beta_{0}$ limit, \eqn{wLLbeta0} and \eqn{wNLLbeta0}, are contained within the RG resummed NLL SCET result.

Finally, the renormalon in the weight function suggests that the dominant contribution to its perturbative expansion is from non-logarithmic terms. We can investigate this numerically by considering the leading logarithmic expansion away from the $P_\gamma \to \Delta$ end point. Following \cite{HLLP+}, we combine all known terms from \eqn{wAlphaExp} and \eqn{wAlphaSCET}, and take the ratio of the various logarithmic terms.  While this misses the contributions of terms beyond NLL and subleading in $\beta_0$, we can hope that the values below are still indicative of the relative contributions of the various terms.  Taking $m_b=4.8$ GeV, $\alpha_s(m_b)=0.22$ and $\mu_{i}^2/m_b^2\sim(\Delta-P_{\gamma})/m_b = 1/9$ as in \cite{HLLP+} we find the following ratios of the logarithmic terms at each order in $\alpha_s$:
\begin{eqnarray}\label{logratios}
\alpha_s^3:\qquad &&O(\log^3) : O(\log^2) : O(\log^1) : O(\log^0)= 1: 2.1 : 1.8 : -6.0
\nonumber\\
\alpha_s^4:\qquad && O(\log^4) :O(\log^3) : O(\log^2) : O(\log^1) : O(\log^0) =1: 3.5 : 2.9 : 0.4 : -26
\nonumber\\
\alpha_s^5:\qquad
&&O(\log^5):O(\log^4) :O(\log^3) : O(\log^2) : O(\log^1) : O(\log^0)=\nonumber\\
&&\hskip3in 1 : 4.9 : 4.2 : 1.0 : -2.3 : -119.
\end{eqnarray}

From these results, we can make two observations.  First, the renormalon ambiguity in the weight function is reflected in the rapid growth of the non-logarithmic terms, which dominate the perturbative expansion.   However, this bad behaviour of perturbation theory is unphysical:  in a consistent approach to $O(1/m_b)$, the renormalon in the weight function will cancel with a corresponding ambiguity in the definitions of the subleading shape functions.  This cancellation would be manifest if the subleading shape functions were consistently extracted from physical observables, but since they are currently modelled, no such cancellation is manifest.  We will see in the next section that the estimated uncertainty in $|V_{ub}|$ from the factorially growing terms is small compared to other sources of error, so we will not attempt in this paper to absorb the renormalon ambiguity into subleading shape functions.   These results do, however, underscore the fact that separating the bad behaviour of perturbation theory from the $O(1/m_b)$ corrections is not a well-defined procedure.

Second, assuming the pattern in \eqn{logratios} continued to hold beyond the large-$\beta_0$ and NLL terms included here, it indicates that terms which are enhanced by more powers of $\log \mu_i^2/m_b^2\sim \log (1/9)\sim -2$ do not dominate over terms with fewer powers of logarithms.  Since the logarithmic terms do not suffer from renormalon ambiguities, and, therefore, no cancellation against the subleading operators is
expected for these terms, this pattern should not change once subleading operators are consistently included.  Thus, these results support the conclusion of \cite{HLLP+} that fixed-order perturbation theory is more appropriate than a leading-log resummation for the extraction of $|V_{ub}|$ (see also \cite{Melnikov, Misiak08}).

\subsection{Determination of $|V_{ub}|$}

\begin{figure}
\centering 
\includegraphics[width=0.7\textwidth]{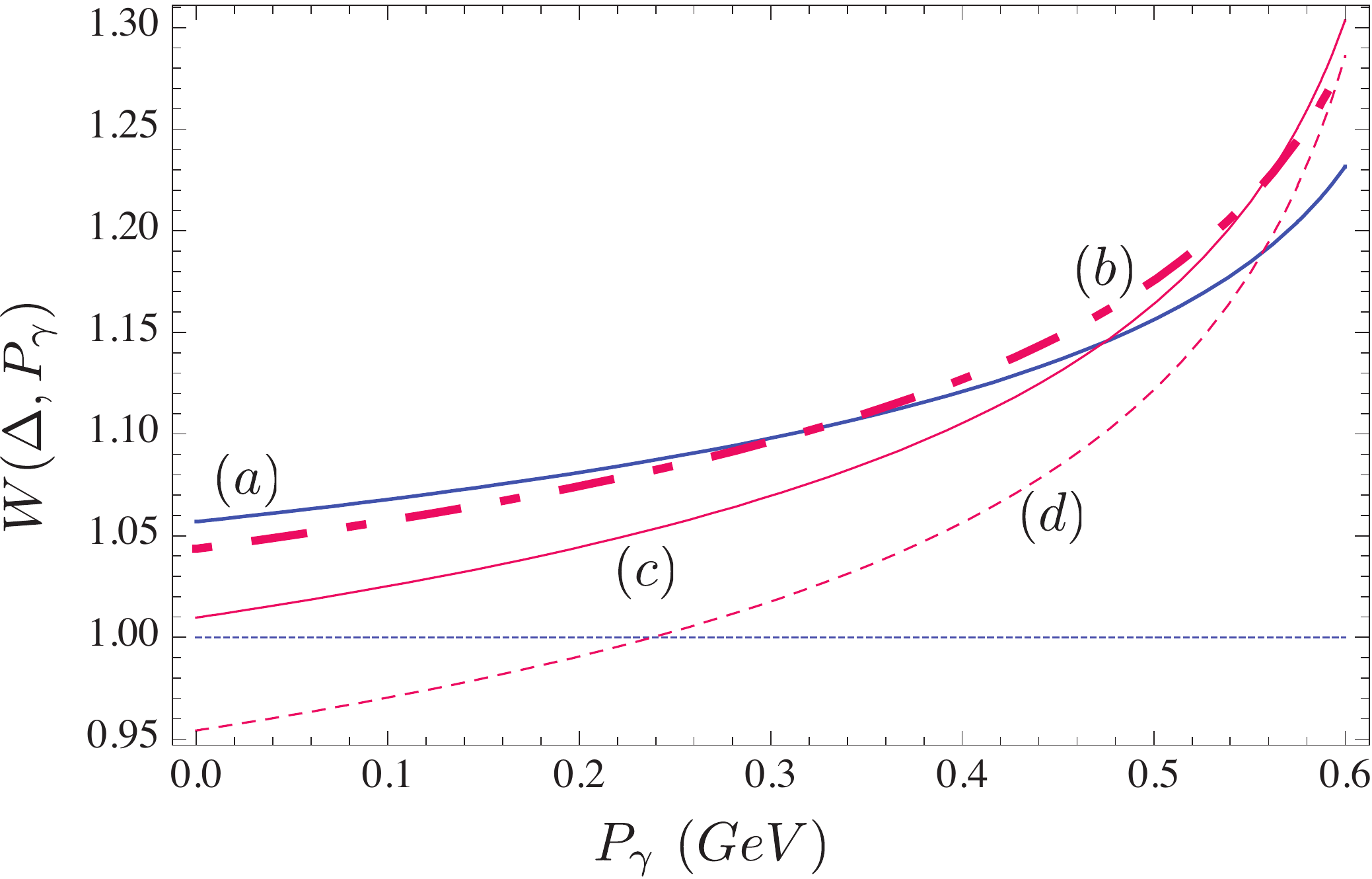}
\caption{$W(\Delta,P_{\gamma})$ from \eqn{wAlphaExp} is shown including terms up to the following order: (a) $O(\alpha_{s}^{2} \beta_{0})$, (b) $O(\alpha_{s}^{3} \beta_{0}^{2})$, (c) $O(\alpha_{s}^{4} \beta_{0}^{3})$ and (d) $O(\alpha_{s}^{5} \beta_{0}^{4})$.}\label{wtAlpha1}
\end{figure}

\begin{figure}
\centering 
\includegraphics[width=0.7\textwidth]{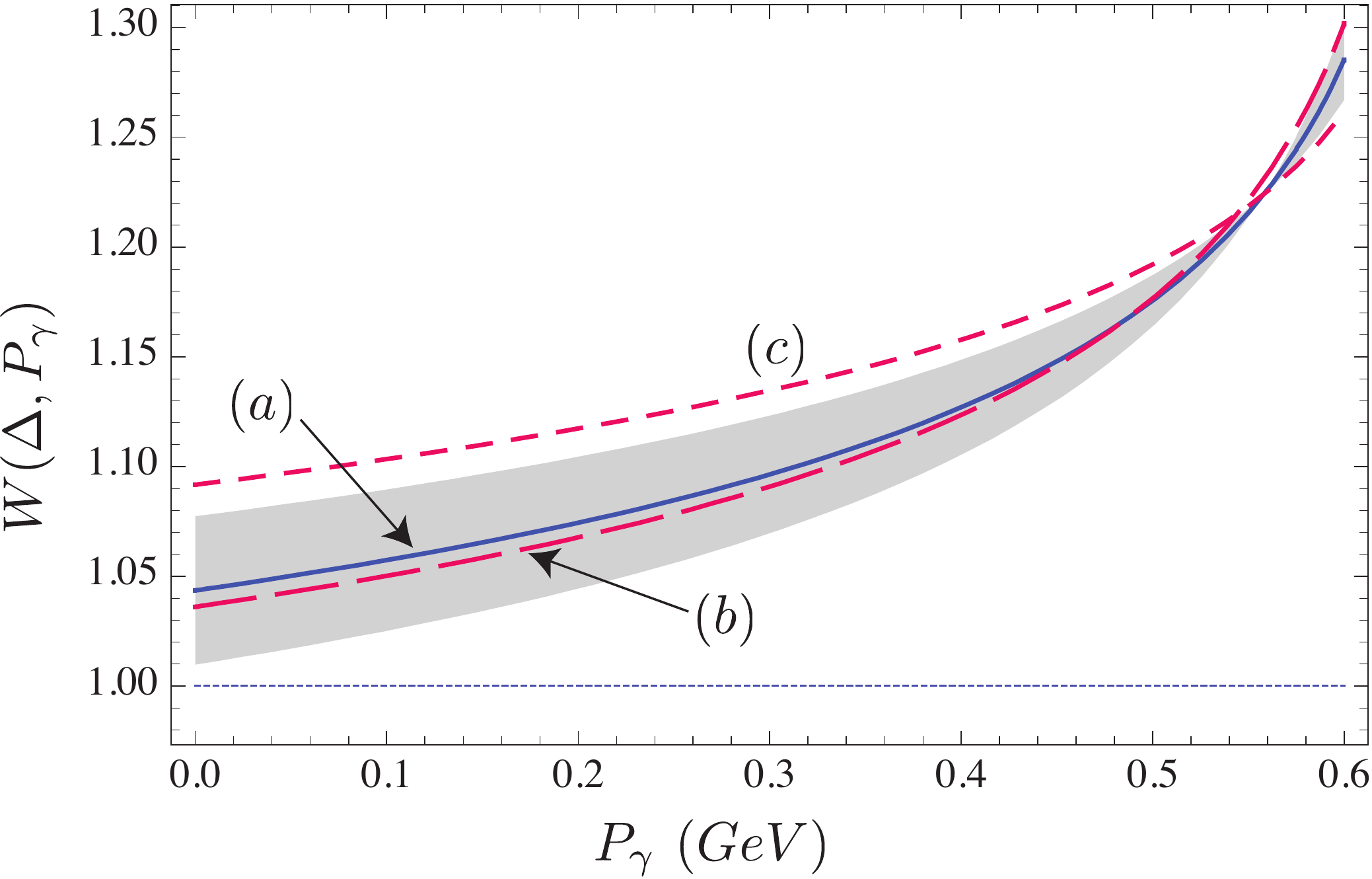}
\caption{(a) $W(\Delta,P_{\gamma})$ with all terms to $O(\alpha_{s}^{3})$ from \eqn{wAlphaExp} and \eqn{wAlphaSCET}. The grey region is the error estimate obtained from the $\alpha_{s}^4 \beta_{0}^3$ term. (b) $W(\Delta,P_{\gamma})$ up to $O(\alpha_{s}^3 \beta^2_0)$ from \eqn{wAlphaExp}. (c) The resummed NLL SCET result, $W(\Delta,P_{\gamma})^{\rm NLL}_{\rm SCET}$.}\label{wtAlpha3}
\end{figure}

From a phenomenological perspective, our results are most useful as an
estimate of the size of higher-order perturbative corrections to the
extraction of $|V_{ub}|$ via \eqn{weightRelation}.  The perturbative
results in \eqn{wAlphaExp} for $\wdp$ are plotted in \fig{wtAlpha1} at
different orders in the  $\alpha_{s}^{n} \beta_{0}^{n-1} $ expansion.
Throughout this section, we will use the values $m_b = 4.8$ GeV and
$\alpha_{s}(m_b) = 0.22$ for numerical evaluations, and take $\Delta =
{m_D^2/m_B}=0.66$ GeV, corresponding to the kinematic cut which removes
the $B\to X_c$ background.  At tree level, the weight function is 1 (the
dotted line in \fig{wtAlpha1} and \fig{wtAlpha3}).  Curve (a) in \fig{wtAlpha1} shows the weight function up to $O(\alpha_{s}^{2} \beta_{0})$, calculated previously in \cite{HLLP+}, while curves (b), (c) and (d) show the results to $O(\alpha_s^3\beta_0^2)$, $O(\alpha_s^4\beta_0^3)$ and $O(\alpha_s^5\beta_0^4)$.   

It is clear from the plots that the perturbative series for $\wdp$ is not converging well, as was discussed in the previous section, due largely to the factorial growth of the constant terms in $W(\Delta, P_\gamma)$.  As we will discuss shortly, the results suggest that the optimal perturbative estimate is obtained by truncating the series at $O(\alpha_s^3)$, and using the $O(\alpha_s^4)$ result as an estimate of the corresponding perturbative uncertainty.  In \fig{wtAlpha3} we therefore compare the fixed-order $\alpha_s^3\beta_0^2$ result to other perturbative estimates of the weight function.  Curve (a) shows all known terms up to $O(\alpha_s^3)$:  the complete NLL terms from \eqn{wAlphaSCET}, combined with the additional large $\beta_0$ terms in \eqn{wAlphaExp} that are higher order in the leading log expansion.   The gray band around the curve gives the perturbative error estimate given by the $O(\alpha_s^4\beta_0^3)$ term.  The result is very close to the large-$\beta_0$ calculation up to $O(\alpha_s^3\beta_0^2)$, shown in Curve (b).  Curve (c) shows the complete NLL resummed result.   

As discussed in the previous section, the integral in \eqn{weightRelation} has a worse perturbative expansion than the weight function itself, since at higher orders in perturbation theory $W(\Delta, P_\gamma)$ is more singular at the endpoint of integration.  Hence, to determine the effects of perturbative corrections on the determination of $|V_{ub}|$, we must look at the perturbative expansion of \eqn{weightRelation} rather than that of $W(\Delta, P_\gamma)$.  For the purposes of estimating the size of higher order terms, we adopt the simple model of the  normalized $B\rightarrow X_{s}\gamma$ spectrum, \eqn{normPhoton}. We obtain $\widehat{\Gamma}_{u}(\Delta)$, the integrated $B\rightarrow X_u\ell\bar\nu_{l}$ decay rate normalized to the tree level value, 
\begin{equation}
\widehat{\Gamma}_u(\Delta)={1\over\Gamma_u}\int_0^\Delta dP_+ {d\Gamma_u\over dP_+}
\end{equation}
shown in Table \ref{table1}. We include several more terms than are explicitly shown in \eqn{wAlphaExp} to demonstrate that the series appears to converge up to $O(\alpha_{s}^4 \beta_{0}^3)$ and then begins to diverge.  This suggests that the optimal perturbative result is given by including all terms up to $O(\alpha_{s}^3)$ and using the $O(\alpha_{s}^4)$ contribution to estimate the perturbative uncertainty.   At this stage, our best estimate of this result is obtained by including all known terms up to $O(\alpha_s^3)$ from \eqn{wAlphaExp} and \eqn{wAlphaSCET}, and estimating the uncertainty from the $O(\alpha_s^4\beta_0^3)$ term.  Table \ref{table2} gives $\widehat{\Gamma}_u (\Delta)$ obtained from the renormalization group resummed LL and NLL weight function in SCET, as well as all terms up to $O(\alpha_{s}^3)$ from \eqn{wAlphaExp} and \eqn{wAlphaSCET}. We see that the NLL result is in agreement with the optimal perturbative value, within the error. The perturbative uncertainty in $|V_{ub}|$, estimated from the $O(\alpha_{s}^4 \beta_{0}^3)$ terms is approximately $0.5\%$, which is far smaller than the order $5 \%$ theoretical uncertainty in $|V_{ub}|$  from subleading shape functions, error in the $b$ quark mass and other sources \cite{neub2loop}.

\begin{table}[b]
\begin{tabular}{c|c|c|c|c|c|c|c}
Tree & $O(\alpha_{s})$ & $O(\alpha_{s}^2 \beta_{0})$& $O(\alpha_{s}^3
\beta_{0}^2)$& $O(\alpha_{s}^4 \beta_{0}^3)$& $O(\alpha_{s}^5
\beta_{0}^4)$ & $O(\alpha_{s}^6 \beta_{0}^5)$ & $O(\alpha_{s}^7
\beta_{0}^6)$ \nonumber \\
\hline
1 & 1.08 & 1.15 & 1.17 & 1.16 &1.12& 1.04 & 0.88  \nonumber \\
\end{tabular}
  \caption{$\widehat{\Gamma}_{u}(\Delta)$ for different orders in the ``large-$\beta_0$" expansion of $W(\Delta,P_{\gamma})$, \eqn{wAlphaExp}.}\label{table1}
\end{table}

\begin{table}
\begin{tabular}{c|c|c|c}
Tree & SCET LL & SCET NLL & All known terms to $O(\alpha_{s}^3)$ \nonumber \\
\hline
1 & 1.10 & 1.18 & 1.17  \nonumber \\
\end{tabular}
\caption{$\widehat{\Gamma}_{u}(\Delta)$ for the resummed LL and NLL weight function in SCET, and all terms up to $O(\alpha_{s}^3)$ from \eqn{wAlphaExp} and \eqn{wAlphaSCET}.}\label{table2}
 \end{table}
 
\section{Conclusions}

We have calculated the Borel transform of the $B\to X_u\ell\bar\nu_\ell$ $P_{+}$ spectrum and $B\rightarrow X_{s}\gamma$ $P_{\gamma}$ spectrum to leading order in $\Lambda_{QCD}/m_b$, from which we determine the Borel transform of the weight function. The leading renormalon in $W(\Delta,P_{\gamma})$ is confirmed to be at $u=1/2$, corresponding to nonperturbative corrections at $O(\Lambda_{QCD}/m_b)$. The $\alpha_{s}^{n} \beta_{0}^{n-1}$ terms are easily obtained from the Borel transform of the weight function and are given analytically to $n=5$. We are able to resum logarithms of $(\Delta-P_{\gamma})/m_b$ in the large $\beta_{0}$ limit of the weight function since the relevant terms in $B[W(\Delta,P_{\gamma})](u)$ are renormalon free. However we show that integrating these terms over $P_\gamma$ introduces a renormalon. Comparing all known terms in the perturbative expansion of the weight function, we find no numerical enhancement of leading logarithms, suggesting that fixed-order perturbation theory is more appropriate than a leading-log resummation.  

From our results we estimate the size of higher-order perturbative corrections on the extraction of $|V_{ub}|$ using a model for the $B\rightarrow X_{s}\gamma$ photon spectrum. We have shown that $\widehat{\Gamma}_{u}(\Delta)$ begins to diverge beyond $O(\alpha_{s}^4 \beta_{0}^3)$ in the $\beta_{0}$ expansion of the weight function. This suggests that the best perturbative estimate is given by including terms up to $O(\alpha_{s}^3)$ with the theoretical uncertainty given by the $\alpha_{s}^4 \beta_{0}^3$ term. We show that this result is in good agreement with the resummed NLL SCET result.  
\section{Acknowledgments}
F.C. acknowledges support from a postdoctoral fellowship of the Generalitat
Valenciana (Beca Postdoctoral d\'{}Excel$\cdot$l\`encia). M.L and S.Z are
supported by the Natural Sciences and Engineering Research Council of Canada.

\appendix
\section{Expanding out the functions}\label{appendixa}

\begin{eqnarray}\label{btosalphacoeff}
C^{s}_{1}(\Delta)&=& -2 \ln ^2\frac{\Delta}{m_b} -7 \ln \frac{\Delta}{m_b} -\frac{4 \pi ^2}{3}-5  \nonumber \\
C^{s}_{2}(\Delta)&= &2 \ln ^3\frac{\Delta}{m_b} +\frac{13}{6}  \ln ^2\frac{\Delta}{m_b}+\left(-\frac{85}{6}+\frac{2 \pi ^2}{3}\right) \ln \frac{\Delta}{m_b} -4 \zeta (3)-\frac{91 \pi ^2}{18}-\frac{631}{36} \nonumber \\
C^{s}_{3}(\Delta)&=-&\frac{7}{3} \ln ^4\frac{\Delta }{m_b}+\frac{1}{3} \ln ^3\frac{\Delta}{m_b}+\left(\frac{275}{18}-\frac{2 \pi ^2}{3}\right) \ln ^2\frac{\Delta}{m_b}+\frac{1}{18} \left(-581+58 \pi ^2\right) \ln\frac{\Delta}{m_b} \nonumber \\ 
&&+ \frac{1}{324} \left(-12727-6366 \pi ^2-108 \pi ^4-13824 \zeta (3)\right)\nonumber \\
C^{s}_{4}(\Delta)&=&3 \ln ^5\frac{\Delta }{m_b}-\frac{35}{12} \ln ^4\frac{\Delta }{m_b}+\left(-\frac{35}{2}+\frac{2 \pi ^2}{3}\right) \ln ^3\frac{\Delta }{m_b}+\left(\frac{6029}{108}-\frac{29 \pi ^2}{6}\right) \ln ^2\frac{\Delta }{m_b} \nonumber \\
&&+ \left(-\frac{9557}{108}+\frac{235 \pi^2}{18}-\frac{\pi^4}{5} \right)\ln \frac{\Delta }{m_b}-72 \zeta (5)-\frac{555}{2}\zeta (3)+\pi ^2
   \left(-\frac{24959}{324}-8 \zeta (3)\right) \nonumber \\
&& -\frac{57 \pi ^4}{10}-\frac{283555}{2592}\nonumber \\
C^{s}_{5}(\Delta)&=&-\frac{62}{15} \ln ^6\frac{\Delta }{m_b}+\frac{33}{5} \ln ^5\frac{\Delta }{m_b}+\left(\frac{395}{18}-\frac{2 \pi^2}{3}\right)\ln ^4\frac{\Delta }{m_b}+\left(-\frac{2543}{27}+\frac{58 \pi^2}{9}\right)\ln ^3\frac{\Delta }{m_b} \nonumber \\
&&+\left(\frac{32171}{162}-\frac{235 \pi^2}{9}+\frac{2 \pi^4}{5}\right)\ln ^2\frac{\Delta }{m_b}+\left(-\frac{50189}{162}+\frac{4429 \pi^2}{81}-\frac{29 \pi^4}{15}\right)\ln\frac{\Delta }{m_b} \nonumber \\
&&-\frac{7392583}{19440}-\frac{154997 \pi^2}{486}-\frac{3932 \pi^4}{75}-\frac{494 \pi^6}{315}-\frac{7452}{5}\zeta(5)\nonumber\\
&&-\left( \frac{205219}{135}+\frac{496 \pi^2}{3} +96\zeta(3) \right)\zeta(3).
\end{eqnarray}

\begin{eqnarray}\label{btoualphacoeff}
C^{u}_{1}(\Delta)&= &-2 \ln ^2\frac{\Delta}{m_b}-\frac{26}{3} \ln \frac{\Delta}{m_b} -2 \pi^2-\frac{13}{36}  \nonumber \\
C^{u}_{2}(\Delta)&=& 2 \ln ^3\frac{\Delta}{m_b} + 3 \ln ^2\frac{\Delta}{m_b}+\left(-\frac{113}{6}+\frac{2 \pi^2}{3}\right) \ln \frac{\Delta}{m_b} -16 \zeta (3)-\frac{41 \pi ^2}{6}+\frac{1333}{144} \nonumber \\
C^{u}_{3}(\Delta)&=&-\frac{7}{3} \ln ^4\frac{\Delta }{m_b}-\frac{2}{9} \ln ^3\frac{\Delta}{m_b}+\left(\frac{359}{18}-\frac{2 \pi ^2}{3}\right) \ln ^2\frac{\Delta}{m_b}+ \left(-\frac{5045}{108}+\frac{34 \pi^2}{9}\right) \ln\frac{\Delta}{m_b} \nonumber \\ 
&&+\frac{168313}{2592}-\frac{2135 \pi^2}{108}-\frac{7 \pi^4}{5}-98 \zeta(3)\nonumber \\
C^{u}_{4}(\Delta)&=& 3 \ln ^5\frac{\Delta }{m_b}-\frac{5}{2} \ln ^4\frac{\Delta }{m_b}+\frac{1}{6}\left(-133+4 \pi^2\right) \ln ^3\frac{\Delta }{m_b}+\left(\frac{16735}{216}-\frac{17 \pi ^2}{3}\right) \ln ^2\frac{\Delta }{m_b} \nonumber \\
&&+\left(-\frac{180229}{1296}+\frac{319 \pi^2}{18}-\frac{\pi^4}{5} \right)\ln\frac{\Delta }{m_b}-432 \zeta (5)-\left( \frac{1807}{6}+40 \pi^2 \right)\zeta(3) -\frac{13129 \pi^2}{432}\nonumber \\
&&-\frac{79 \pi^4}{6}+\frac{11428313}{31104} \nonumber \\
C^{u}_{5}(\Delta)&=& -\frac{62}{15} \ln ^6\frac{\Delta }{m_b}+\frac{94}{15} \ln ^5\frac{\Delta }{m_b}+\left(\frac{479}{18}-\frac{2 \pi^2}{3} \right)\ln ^4\frac{\Delta }{m_b}+\left(-\frac{2215}{18}+\frac{68 \pi^2}{9} \right)\ln ^3\frac{\Delta }{m_b}\nonumber \\
&&+\left(\frac{21581}{72}-\frac{319 \pi^2}{9}+\frac{2 \pi^4}{5} \right)\ln ^2\frac{\Delta }{m_b} +\left(-\frac{668117}{1296}+\frac{13535 \pi^2}{162}-\frac{34 \pi^4}{15} \right)\ln\frac{\Delta }{m_b}-4920 \zeta (5)\nonumber \\
&&+\left(\frac{29741}{54}-\frac{1408 \pi^2}{3}-672 \zeta(3) \right) \zeta(3)+\frac{8231 \pi^2}{48}-\frac{2774 \pi^6}{315}-\frac{1649 \pi^4}{30}+\frac{64526377}{31104}.
\end{eqnarray}

\begin{eqnarray}\label{wAlphaExp}
W_1(\Delta, P_\gamma)&=&-\frac{5}{3} \ln \frac{\Delta-P_{\gamma}}{m_b}-\frac{2 \pi ^2}{3}+\frac{167}{36}  \nonumber \\
W_2(\Delta, P_\gamma)&=&\frac{5}{6} \ln^2 \frac{\Delta-P_{\gamma}}{m_b} -\frac{14}{3} \ln \frac{\Delta-P_{\gamma}}{m_b} + \frac{3857}{144}-\frac{16 \pi ^2}{9}-12 \zeta (3)\nonumber \\
W_3(\Delta, P_\gamma)&=&-\frac{5}{9} \ln^3 \frac{\Delta-P_{\gamma}}{m_b}+\frac{14}{3} \ln^2 \frac{\Delta-P_{\gamma}}{m_b}+\left(\frac{5 \pi^{2}}{9}-\frac{1559}{108}\right) \ln\frac{\Delta-P_{\gamma}}{m_b}+ \frac{90043}{864}\nonumber \\
&&-\frac{13 \pi ^2}{108}-\frac{16 \pi^4}{15}-\frac{166}{3} \zeta (3) \nonumber\\
W_4(\Delta, P_\gamma)&=&\frac{5}{12}\ln^4 \frac{\Delta-P_{\gamma}}{m_b}-\frac{14}{3}\ln^3 \frac{\Delta-P_{\gamma}}{m_b}+\left(\frac{1559}{72}-\frac{5 \pi^{2}}{6}\right) \ln^2 \frac{\Delta-P_{\gamma}}{m_b}\nonumber \\
&&+\left(\frac{14 \pi^{2}}{3}-\frac{65545}{1296}\right)\ln \frac{\Delta-P_{\gamma}}{m_b}-360 \zeta(5)-\left(\frac{71}{3}+32 \pi^2 \right)\zeta(3)\nonumber \\
&&-\frac{112 \pi^4}{15}+\frac{60449 \pi^2}{1296}+\frac{14830973}{31104} \nonumber\\
W_5(\Delta, P_\gamma)&=&-\frac{1}{3}\ln^5 \frac{\Delta-P_{\gamma}}{m_b}+\frac{14}{3} \ln^4 \frac{\Delta-P_{\gamma}}{m_b}+\left(-\frac{1559}{54}+\frac{10 \pi^2}{9} \right) \ln^3 \frac{\Delta-P_{\gamma}}{m_b}\nonumber \\
&& +\left(\frac{65545}{648}-\frac{28 \pi^2}{3} \right) \ln^2 \frac{\Delta-P_{\gamma}}{m_b}+\left(-\frac{266605}{1296}+\frac{1559 \pi^2}{54}-\frac{\pi^4}{3} \right) \ln \frac{\Delta-P_{\gamma}}{m_b}\nonumber \\
&& -\frac{17148}{5}\zeta(5)+\left(-576 \zeta(3)+\frac{20709}{10}-304\pi^2\right)\zeta(3)-\frac{152 \pi^6}{21} -\frac{127 \pi^4}{50}\nonumber \\
&&+\frac{1906687 \pi^2}{3888}+\frac{381772549}{155520}. 
\end{eqnarray}

\section{The Weight Function to NLL Order}\label{appendixb}

The renormalization group resummed NLL weight function has been calculated in SCET, \cite{p+cut,Neubert,BauerMan,neub2loop}.  By expanding $W(\Delta,P_{\gamma})^{\rm NLL}_{\rm SCET}$ in $\alpha_{s}(m_b)$ and re-expanding the logarithms of $\mu_{i}^2/m_b^2$ and  $\mu_{i}^2/(m_b(\Delta-P_{\gamma}))$ we find

\begin{equation}W(\Delta,P_\gamma)^{\rm NLL}_{\rm SCET}=1+\sum_{i=1}^\infty W_n(\Delta, P_\gamma)^{\rm NLL}_{\rm SCET} {\alpha_s(m_b)^n C_F\over (4\pi)^n}
\end{equation}
and the first coefficients are given by:
\begin{eqnarray}\label{wAlphaSCET}
W_1(\Delta, P_\gamma)^{\rm NLL}_{\rm SCET}&=&-\frac{5}{3} \ln \frac{\Delta-P_{\gamma}}{m_b}-\frac{2 \pi ^2}{3}+\frac{167}{36}  \nonumber \\
W_2(\Delta, P_\gamma)^{\rm NLL}_{\rm SCET} &=&\left(\frac{5 \beta_0}{6}+\frac{92}{27} \right) \ln^2 \frac{\Delta-P_\gamma}{m_b}+\left(-\frac{14 \beta_0}{3}+\frac{128}{3} \zeta(3)+\frac{85 \pi^2}{27}-\frac{5122}{81}\right) \ln\frac{\Delta-P_\gamma}{m_b}\nonumber \\
W_3(\Delta, P_\gamma)^{\rm NLL}_{\rm SCET}&=&\left(-\frac{5 \beta^2_0}{9}-\frac{92 \beta_0}{27}-\frac{1616}{243}\right) \ln^3 \frac{\Delta-P_\gamma}{m_b}+\left(\frac{14\beta^2_0}{3}+\left(-\frac{64}{3}\zeta(3)-\frac{65 \pi^2}{27}\right.\right. \nonumber \\
&& \left. \left.+\frac{11501}{162}\right) \beta_0-\frac{2560}{27} \zeta(3)-\frac{512 \pi^4}{135}-\frac{2392 \pi^2}{243}+\frac{68155}{162}\right) \ln^2\frac{\Delta-P_\gamma}{m_b}\nonumber \\
W_4(\Delta, P_\gamma)^{\rm NLL}_{\rm SCET}&=&\left(\frac{5 \beta^3_0}{12}+\frac{253 \beta^2_0}{81}+\frac{808 \beta_0}{81}+\frac{27584}{2187}\right) \ln^4 \frac{\Delta-P_\gamma}{m_b} +\left(-\frac{14 \beta^3_0}{3}+\left( \frac{128}{9}\zeta(3) \right.\right.  \nonumber \\
&&  \left. +\frac{175 \pi^2}{81}-\frac{19981}{243}\right) \beta^2_0+\left(\frac{2560}{27}\zeta(3)+\frac{512 \pi^4}{135}+\frac{3220 \pi^2}{243}-\frac{243991}{486}\right) \beta_0 \nonumber\\
&&\left. +\frac{65536}{27}\zeta(5)+\frac{47104}{243}\zeta(3)+\frac{2048 \pi^4}{243}+\frac{56560 \pi^2}{2187}-\frac{21384356}{6561}\right)\ln^3\frac{\Delta-P_\gamma}{m_b}\nonumber \\
W_5(\Delta, P_\gamma)^{\rm NLL}_{\rm SCET}&=&\left(-\frac{\beta^4_0}{3}-\frac{230 \beta^3_0}{81}-\frac{2828 \beta^2_0}{243}-\frac{55168 \beta_0}{2187}-\frac{462080}{19683} \right) \ln^5\frac{\Delta-P_\gamma}{m_b}\nonumber \\
&&+\left(\frac{14 \beta^4_0}{3}+\left(-\frac{32}{3}\zeta(3)-\frac{55 \pi^2}{27}+\frac{89585}{972}\right)\beta^3_0+\left(-\frac{7040}{81}\zeta(3)-\frac{1408 \pi^4}{405}\right.\right. \nonumber\\
&&\left.-\frac{11132 \pi^2}{729}+\frac{548459}{972}\right)\beta^2_0+\left(-\frac{32768}{9}\zeta(5)-\frac{23552}{81}\zeta(3)-\frac{1024 \pi^4}{81} \right.\nonumber \\
&&\left.-\frac{35552 \pi^2}{729}+\frac{3675094}{729}\right) \beta_0-\frac{1310720}{243}\zeta(5)-\frac{827392}{2187}\zeta(3)-\frac{262144 \pi^6}{15309} \nonumber \\
&&\left.-\frac{188416 \pi^4}{10935}-\frac{1213696\pi^2}{19683}+\frac{1371073480}{59049}\right)\ln^4\frac{\Delta-P_\gamma}{m_b}.
\end{eqnarray}
We verify that the leading $\beta_0$ terms agree with \eqn{wAlphaExp}.


\begin{thebibliography}{99} 

\bibitem{OPE}
 J.~Chay, H.~Georgi and B.~Grinstein,
  Phys.\ Lett.\  B {\bf 247} (1990) 399;\\
M.~A.~Shifman and M.~B.~Voloshin,
  Sov.\ J.\ Nucl.\ Phys.\  {\bf 41} (1985) 120; \\
 I.~I.~Y.~Bigi, N.~G.~Uraltsev and A.~I.~Vainshtein,
  Phys.\ Lett.\  B {\bf 293} (1992) 430  [Erratum-ibid.\  B {\bf 297} (1993) 477]  [hep-ph/9207214]; \\
 I.~I.~Y.~Bigi, M.~A.~Shifman, N.~G.~Uraltsev and A.~I.~Vainshtein,
  Phys.\ Rev.\ Lett.\  {\bf 71} (1993) 496 [hep-ph/9304225];\\
 A.~V.~Manohar and M.~B.~Wise,
  Phys.\ Rev.\  D {\bf 49} (1994) 1310 [hep-ph/9308246].



\bibitem{mXcut}
A.~F.~Falk, Z.~Ligeti and M.~B.~Wise,
  Phys.\ Lett.\  B {\bf 406} (1997) 225 [hep-ph/9705235];\\
  R.~D.~Dikeman and N.~G.~Uraltsev,
  Nucl.\ Phys.\  B {\bf 509} (1998) 378 [hep-ph/9703437].

\bibitem{p+cut}

S.~W.~Bosch, B.~O.~Lange, M.~Neubert and G.~Paz,
Nucl.\ Phys.\ B {\bf 699} (2004) 335 [hep-ph/0402094];\\
S.~W.~Bosch, B.~O.~Lange, M.~Neubert and G.~Paz,
Phys.\ Rev.\ Lett.\  {\bf 93} (2004) 221801 [hep-ph/0403223].


\bibitem{shapefunct}
I.~I.~Y.~Bigi, M.~A.~Shifman, N.~G.~Uraltsev and A.~I.~Vainshtein,
  Int.\ J.\ Mod.\ Phys.\  A {\bf 9} (1994) 2467 [hep-ph/9312359].

\bibitem{shapefunct2}
M.~Neubert,
Phys.\ Rev.\ D {\bf 49} (1994) 3392 [hep-ph/9311325];\\
  M.~Neubert,
  Phys.\ Rev.\  D {\bf 49} (1994) 4623 [hep-ph/9312311].
  
  \bibitem{Leibovich}
  A.~K.~Leibovich, I.~Low and I.~Z.~Rothstein,
  Phys.\ Rev.\  D {\bf 61} (2000) 053006 [hep-ph/9909404]; \\
  A.~K.~Leibovich, I.~Low and I.~Z.~Rothstein,
  Phys.\ Lett.\  B {\bf 486} (2000) 86 [hep-ph/0005124].

 \bibitem{HLLP+} 
  A.~H.~Hoang, Z.~Ligeti and M.~Luke,
  Phys.\ Rev.\  D {\bf 71} (2005) 093007 [hep-ph/0502134].

\bibitem{subleadingshape1}
C.~W.~Bauer, M.~E.~Luke and T.~Mannel,
  Phys.\ Rev.\  D {\bf 68} (2003) 094001 [hep-ph/0102089];\\
 C.~W.~Bauer, M.~Luke and T.~Mannel,
  Phys.\ Lett.\  B {\bf 543} (2002) 261 [hep-ph/0205150]
  
 \bibitem{subleadingshape2}
  S.~W.~Bosch, M.~Neubert and G.~Paz,
  JHEP {\bf 0411} (2004) 073  [hep-ph/0409115].
  
  \bibitem{neub2loop}
  B.~O.~Lange, M.~Neubert and G.~Paz,
  JHEP {\bf 0510} (2005) 084   [hep-ph/0508178].

\bibitem{Stewart:2004a}
 K.S.M.~Lee and I.M.~Stewart,
 Nucl.\ Phys.\  B {\bf 721} (2005) 325  [hep-ph/0409045].
 
 \bibitem{Beneke:2004}
 M.~Beneke, F.~Campanario, T.~Mannel and B.~Pecjak,
 JHEP {\bf 06} (2005) 071 [hep-ph/0411395].	

\bibitem{Lee:2006wn}
  S.~J.~Lee, M.~Neubert and G.~Paz,
  Phys.\ Rev.\  D {\bf 75} (2007) 114005  [hep-ph/0609224].
  
\bibitem{Lange}
B.~O.~Lange,
  JHEP {\bf 0601} (2006) 104  [hep-ph/0511098].

\bibitem{BB}
  M.~Beneke and V.~M.~Braun,
 [hep-ph/0010208].
  
 \bibitem{Ball} 
 P.~Ball, M.~Beneke and V.~M.~Braun,
  Nucl.\ Phys.\  B {\bf 452} (1995) 563 [hep-ph/9502300].
 
\bibitem{thooft} 
G.~'t Hooft, in {\it The Whys of Subnuclear Physics},
edited by A.~Zichichi (Plenum, New York, 1978).

\bibitem{GardiAndersen} 
 J.~R.~Andersen and E.~Gardi,
 JHEP {\bf 0701} (2007) 029  [hep-ph/0609250]; \\
  E.~Gardi,
  JHEP {\bf 0404} (2004) 049 [hep-ph/0403249].

\bibitem{Gambino} 
P.~Gambino, E.~Gardi and G.~Ridolfi,
  JHEP {\bf 0612} (2006) 036  [hep-ph/0610140];\\
 V.~Aquila, P.~Gambino, G.~Ridolfi and N.~Uraltsev,
  Nucl.\ Phys.\  B {\bf 719} (2005) 77  [hep-ph/0503083].
  
\bibitem{Neubert}
  M.~Neubert,
  Eur.\ Phys.\ J.\  C {\bf 40} (2005) 165 [hep-ph/0408179].

\bibitem{BauerMan}
  C.~W.~Bauer and A.~V.~Manohar,
  Phys.\ Rev.\  D {\bf 70} (2004) 034024 [hep-ph/0312109].
 
 \bibitem{MLphoton}
  Z.~Ligeti, M.~E.~Luke, A.~V.~Manohar and M.~B.~Wise,
  Phys.\ Rev.\  D {\bf 60} (1999) 034019 [hep-ph/9903305].
 
 \bibitem{ARF1} 
  U.~Aglietti, G.~Ricciardi and G.~Ferrera,
  Phys.\ Rev.\  D {\bf 74} (2006) 034004 [hep-ph/0507285]; \\
 U.~Aglietti, G.~Ricciardi and G.~Ferrera,
  Phys.\ Rev.\  D {\bf 74} (2006) 034006 [hep-ph/0509271].


\bibitem{Melnikov}
  K.~Melnikov and A.~Mitov,
  Phys.\ Lett.\  B {\bf 620} (2005) 69  [hep-ph/0505097].

\bibitem{Misiak08}
  M.~Misiak,
  arXiv:0808.3134 [hep-ph].
\end{thebibliography}
\end{document}